\documentclass[9pt,conference]{IEEEtran}

\usepackage{multirow}
\usepackage{amsmath}
\usepackage{float}
\usepackage{subcaption}
\usepackage{graphicx}
\usepackage[utf8]{inputenc} % allow utf-8 input
\usepackage[T1]{fontenc}    % use 8-bit T1 fonts
\usepackage{hyperref}       % hyperlinks
\usepackage{url}            % simple URL typesetting
\usepackage{booktabs}       % professional-quality tables
\usepackage{amsfonts}       % blackboard math symbols
\usepackage{nicefrac}       % compact symbols for 1/2, etc.
\usepackage{microtype}      % microtypography
\usepackage{lipsum}
\usepackage[affil-sl]{authblk}

\title{Digital Low-Level RF control system for Advanced Light Source Storage Ring}

\author{Qiang Du\IEEEauthorrefmark{1}\thanks{QDu@lbl.gov}}
\author{Lawrence Doolittle}
\author{Michael Betz}
\author{Benjamin Flugstad}
\author{Massimiliano Vinco}
\author{Kenneth Baptiste}
\affil{
   Lawrence Berkeley National Laboratory\\
   1 Cyclotron Rd, Berkeley, CA, 94720 USA
}

\begin{document}
\maketitle

\begin{abstract}
    We have commissioned the digital Low Level RF (LLRF) system for storage ring
    RF at Advanced Light Source at Lawrence Berkeley National Lab (LBNL). The
    system is composed of 42 synchronous sampling channels for feedback control,
    diagnostics, and interlocks. The closed loop RF amplitude and phase
    stability is measured as $< 0.1\%$ and $< 0.1 ^\circ$ respectively, and the
    real-time machine protection interlock latency is measured $< 2.5 \mu$s. We
    have also developed PLC-FPGA-EPICS interfaces to support system
    configurations between hybrid operation modes using two klystrons driving
    two RF cavities at 500MHz resonance frequency. The deployed LLRF system has
    been operating since March 2017.
\end{abstract}

\section{Introduction}

The Advanced Light Source (ALS) at Lawrence Berkeley National Laboratory is a
U.S. Department of Energy's synchrotron light source user facility that is
operational since 1993.  With circumference of 196.8 m, the ALS Storage Ring
(SR) keeps electron beam current of 500 mA at 1.9 GeV under multi--bunch mode
user operation to deliver synchrotron X-rays to surrounding 40 experimental end
stations.  As the electron beam loses energy every turn due to synchrotron
radiation, two normal conducting RF cavities provides a total of $\sim 1.3$ MV
acceleration voltage and keeps the beam energy constant against variable beam
loads and many sources of instabilities.

\begin{figure}[H]
    \centering
    \includegraphics[angle=-90,width=\linewidth]{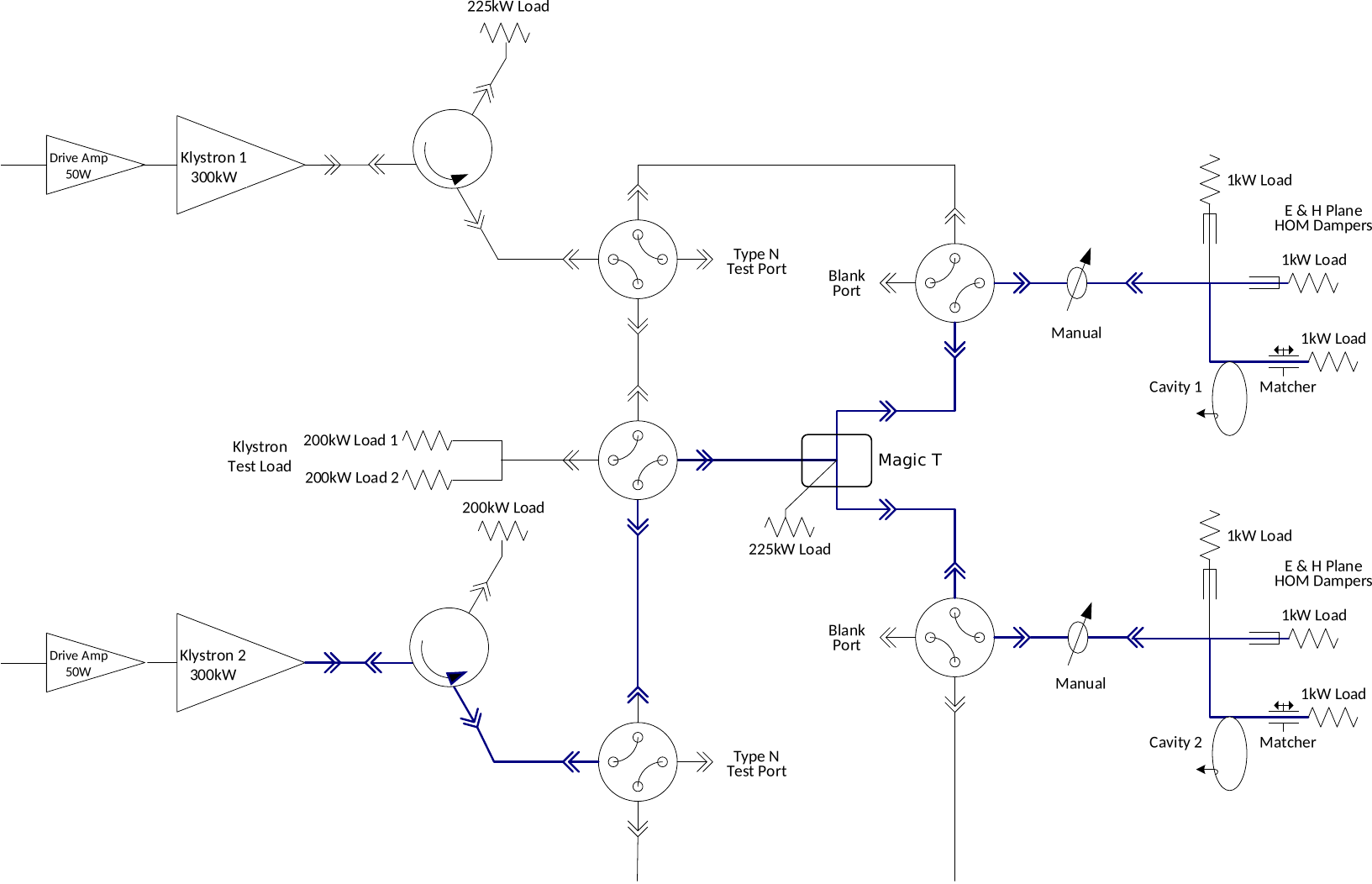}
    \caption{Configurable waveguide mode to drive two cavities}
    \label{fig:wg_diagram}
\end{figure}

The two cavities are driven by two 300 kW klystrons at 499.642 MHz through a
waveguide matrix system, which can be configured to switch RF drive mode among
different modes.
As shown in Figure  \ref{fig:wg_diagram}, 5 configurable RF drive modes includes
each klystron driving one cavity, one klystron driving two cavities or test
loads for RF test.

Table \ref{tab:rf_parameters} shows typical values of storage ring (SR) RF power
requirement for nominal user operation for both ALS at 1.9 Gev 500 mA multi
bunch mode, and the planned ALS-Upgrade (ALS-U) project at 2.0 GeV.

\begin{table}[H]
    \centering
    \begin{tabular}{lccc}
      \toprule
                            & ALS    & ALS-U  & \\
      \midrule
        Number of Klystrons & \multicolumn{2}{c}{2} & \\
        Number of Cavities  & \multicolumn{2}{c}{2} & \\
        Cavity $Q_0$        & \multicolumn{2}{c}{28670} & \\
        Cavity $Q_L$        & \multicolumn{2}{c}{16700} & \\
        Harmonic number $h$ & 328       & 327       & \\
        Circumference       & 196.8     & 195.94    & m \\
        Beam energy         & 1.9       & 2.0       & GeV   \\
        Cavity RF Frequency & 499.64    & 500.394   & MHz   \\
        $\frac{R}{Q}$ (ea)  & 4.9       & 4.9       & M$\Omega$\\
        Cavity voltage      & 671       & 300       & kV    \\
        $\beta$             & 2.9       & 10.07     &       \\
        Energy loss per trun& 329       & 329       & keV   \\
        BM Beam Power       & 141       & 125       & kW    \\
        ID Beam Power       & 42        & 35        & kW    \\
        3HC Beam Power      & 7.3       & 4.4       & kW    \\
        Parasitic Beam Power& 2.9 (est.)& 2.2 (est.)& kW    \\
        Total Beam Power    & 192.9     & 166.9     & kW    \\
        Cavity Power(no beam)&46        & 9.2       & kW    \\
        Cavity Power(beam)  & 142.5     & 127.6     & kW    \\
        Waveguide Loss      & 9.2 (est.)& 2.6 (est.)& kW    \\
        High Power Amplifier& 294.0     & 257.8     & kW    \\
      \bottomrule
    \end{tabular}
    \caption{ALS and ALS-U Storage Ring RF parameters}
    \label{tab:rf_parameters}
\end{table}

As part of the ALS RF system upgrade project, the digital LLRF system was
designed to replace the analog controller \cite{lo1995amplitude} and
corresponding RF interlock systems, with design parameters and requirements
listed in Table \ref{tab:llrf_requirement}.  The digital LLRF system was
installed and commissioned in March 2017, and was operational since then.  It is
expected that the current LLRF design would still meet ALS-U SR RF
specifications, except ALS-U would have additional LLRF control requirements to
its accumulator ring RF.

\begin{table}[H]
    \centering
      \begin{tabular}{lccc}
          \toprule
                                & Analog LLRF & Digital LLRF & \\
          \midrule
          Num. RF Drive Modes   & 1           & 5            & \\
          Num. RF Signals       & 4           & 42           & \\
          Amp. Loop Bandwidth   & 3.5         & 1            & kHz \\
          Phase Loop Bandwidth  & 3.6         & 1            & kHz \\
          Phase Jitter [10Hz,1MHz] & $< 3$    & $< 0.6$      & ps \\
          Interlock latency     & $< 100$     & $< 4$        & $\mu$s\\
          Amp. stability        & 0.56        & $< 0.1$      & \%    \\
          Phase stability       & 1.8         & $< 0.1$      & $^\circ$\\
          \bottomrule
      \end{tabular}
      \caption{Analog LLRF performance and requirements to digital LLRF control}
      \label{tab:llrf_requirement}
\end{table}

\begin{figure}[H]
    \centering
    \includegraphics[page=1, width=\linewidth]{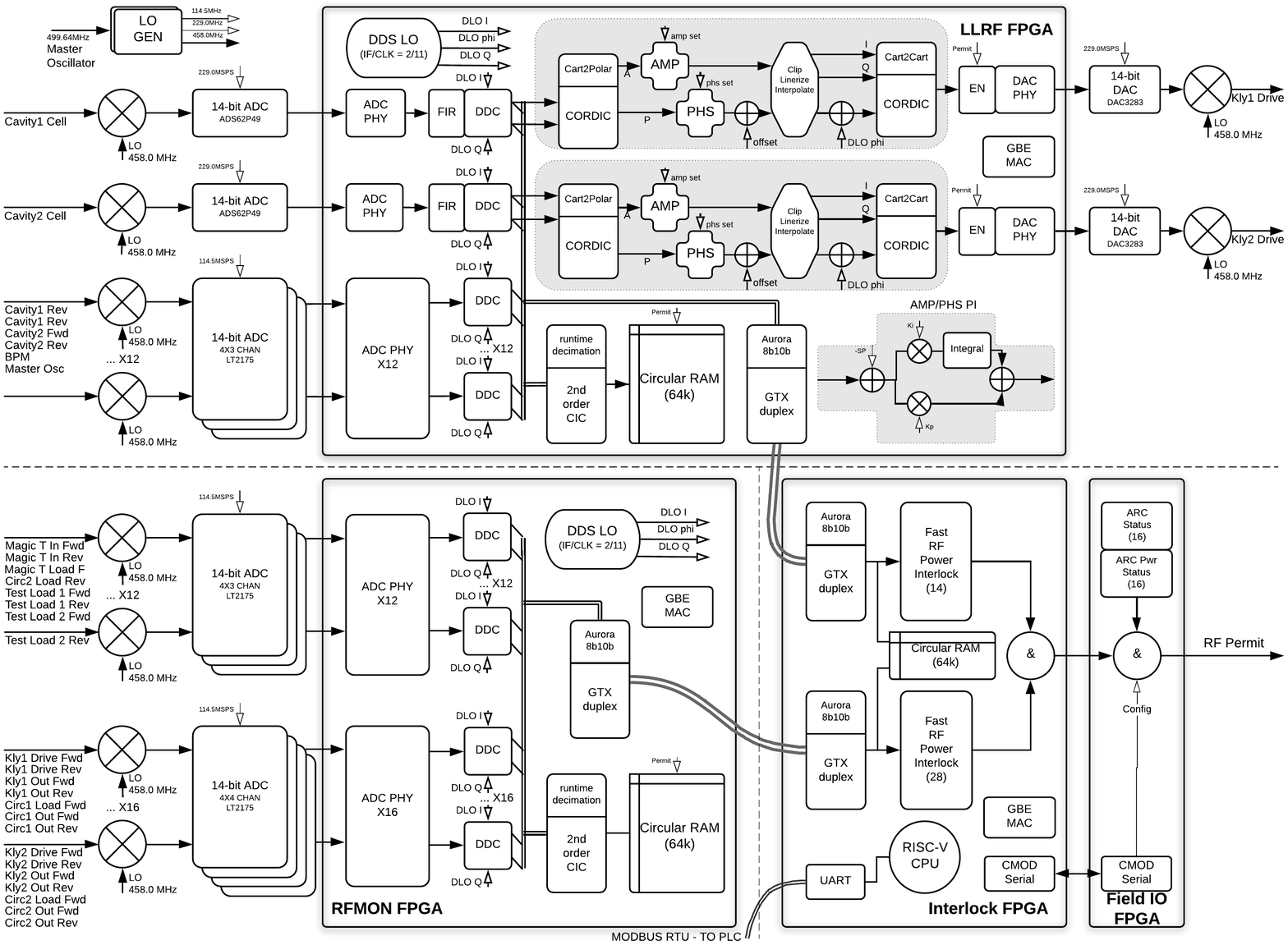}
    \caption{ALS storage ring digital LLRF system overview}
    \label{fig:llrf_diagram}
\end{figure}

The digital LLRF consists of three connected FPGA chassis: LLRF chassis, RF
monitor (RFMON) chassis and Fast Interlock chassis as shown in
\ref{fig:llrf_diagram}. The LLRF chassis generates a common LO reference from
ALS master oscillator for synchronous digitizing and digital signal processing
clocks for all 42 RF signals, and two processed IQ streams from LLRF and RFMON
chassis are sent to Fast Interlock chassis for centralized interlock processing
together with 16 channels of ARC detectors.

Each FPGA chassis has Gigabit Ethernet communication to a local computer that
hosts EPICS Input-Output-Controller(IOC) to access all raw and processed
registers, and configurable waveforms.
Fast Interlock chassis has a dedicated communication to master interlock PLC, so
that all interlock settings are directly available from PLC, which holds the RF
drive mode and waveguide matrix configuration information.

The PLC-FPGA interlock system is designed to be software--free, so that the
operation reliability and interlock invulnerability would not be impacted by 
external components such as EPICS, network or any operating system issues.

\section{Hardware design}
\label{sec:hardware}

\subsection{Frequency configuration}

    Both LO and sampling frequency are derived from ALS master oscillator $f_\text{MO}$.
    \begin{align*}
        f_\text{MO} &= 499.645\,\text{MHz}\\
        f_\text{LO} &= \frac{11}{12} \cdot f_\text{MO} = 458.008\,\text{MHz}\\
        f_\text{IF} &= \frac{ 1}{12} \cdot f_\text{MO} = 41.636\,\text{MHz} \\
        f_\text{S1} &= \frac{1}{2}   \cdot f_\text{LO} = 229.004\,\text{MHz} = f_\text{dsp}\\
        f_\text{S2} &= \frac{1}{4}   \cdot f_\text{LO} = 114.502\,\text{MHz}
    \end{align*}

\subsection{Low-Level RF chassis}

The LLRF chassis uses Abaco Systems FMC150 and FMC112 with Xilinx KC705 FPGA
carrier board for digital platform.  FMC150 provides 2 14-bits ADC and 2 16-bits
DAC channels at $f_\text{S1}$ for precision control, and FMC112 provides 12
14-bits ADC at $f_\text{S2}$ for RF monitoring.
Two Gigabit transceivers (GTX) links are used for inter--FPGA communication via optic
fiber (SFP+) for interlocking and timing distribution respectively.

\begin{figure}[H]
    \centering
    \includegraphics[width=\linewidth]{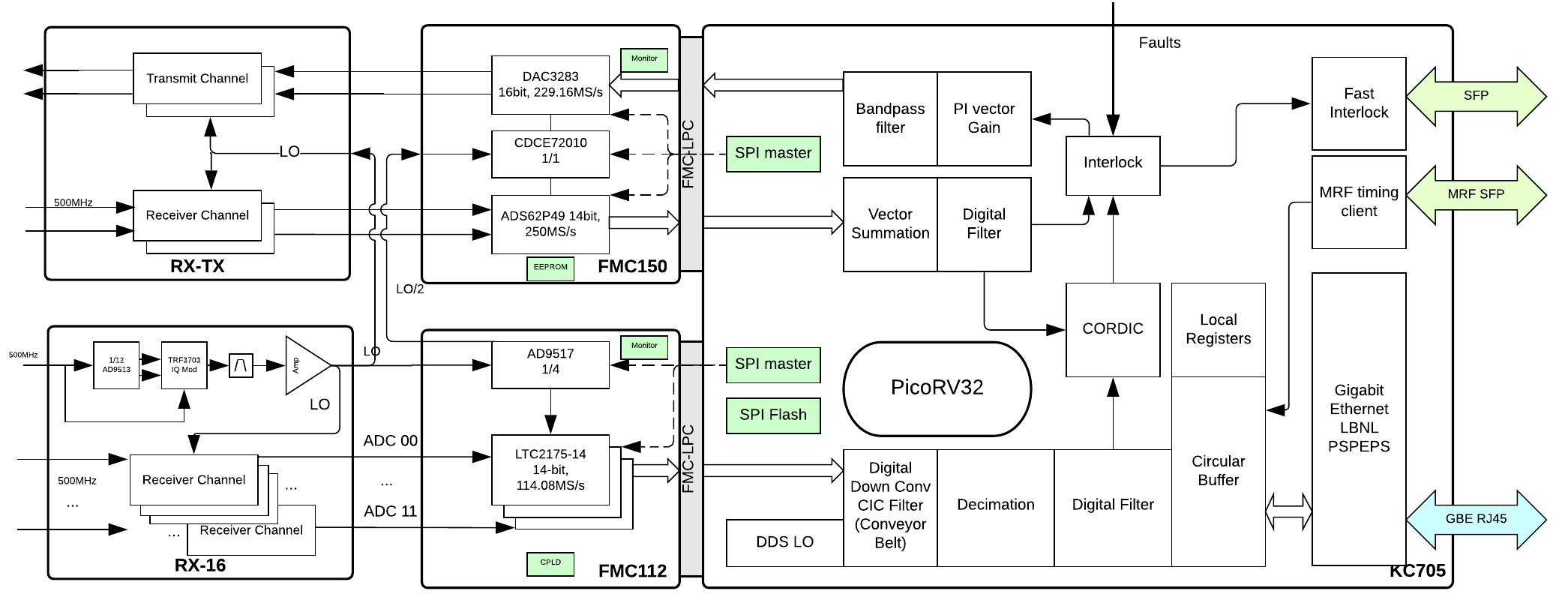}
    \caption{LLRF chassis hardware configuration}
    \label{fig:llrf_hardware}
\end{figure}

% \begin{figure}[H]
%   \begin{subfigure}{\linewidth}
%      % \includegraphics[width=\linewidth]{fig/llrf_chassis_conf.pdf}
%       \includegraphics[width=\linewidth]{fig/llrf_chassis.pdf}
%       \caption{LLRF chassis hardware configuration}
%   \end{subfigure}
%   \begin{subfigure}{.6\linewidth}
%       \includegraphics[width=\linewidth]{fig/llrf_chassis_photo.jpg}
%       \caption{LLRF chassis in production}
%   \end{subfigure}
% \end{figure}

\subsubsection{Single Side Band LO generation}

\begin{figure}[H]
  \begin{subfigure}{0.45\linewidth}
      % \includegraphics[width=\linewidth]{fig/llrf_mo_PN.pdf}
      % \caption{External MO (R\&S SMA100) Phase Noise}
      \includegraphics[width=\linewidth]{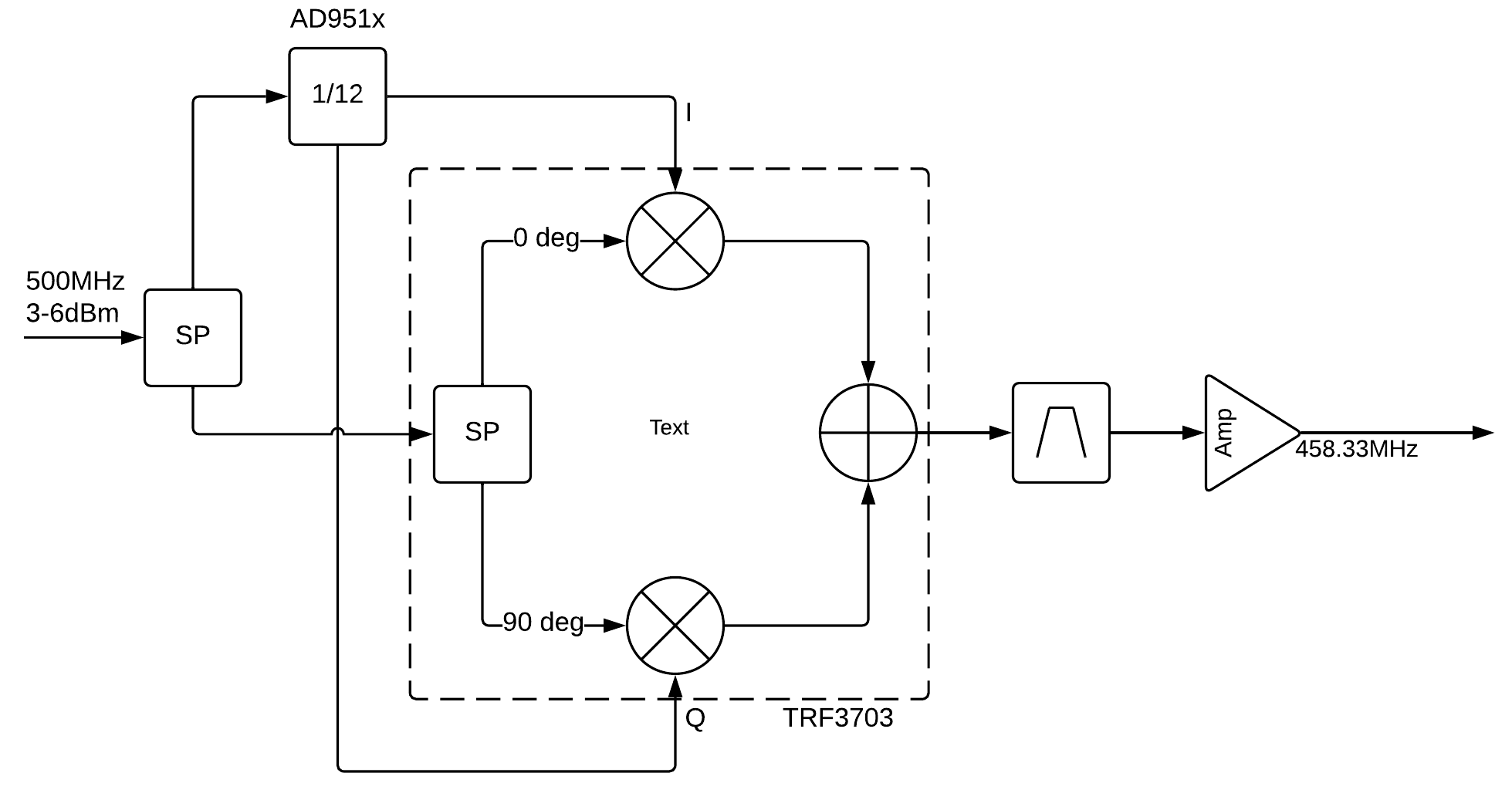}
      \caption{SSB LO generation scheme}
      \label{fig:lo_gen}
  \end{subfigure}
  \begin{subfigure}{0.45\linewidth}
      \includegraphics[width=\linewidth]{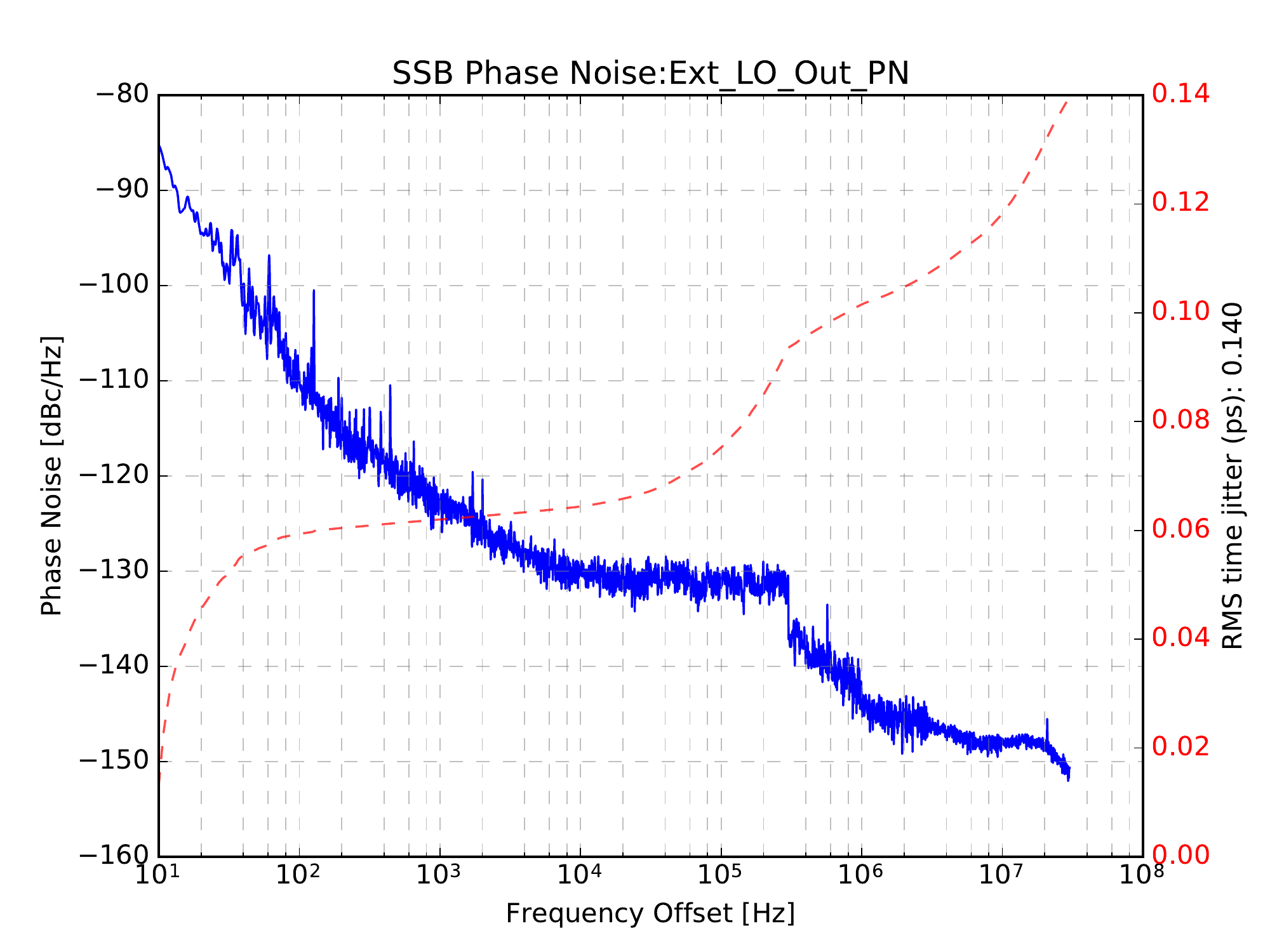}
      \caption{Phase Noise of Generated LO}
      \label{fig:lo_phase_noise}
  \end{subfigure}
\end{figure}

$f_\text{LO}$ is generated using single side band modulation by a frequency
divider AD9513 and a vector modulator TRF3703, as shown in Figure
\ref{fig:lo_gen}. Because both chips are not sensitive to clock signal level,
the generated LO level is stable against MO signal level variations. The
carrier feed through and image frequency suppression is also benefited as shown
in Table \ref{table:lo_gen}. When using an external signal source with 107 fs
rms phase jitter at $f_\text{MO}$, the measured phase noise of $f_\text{LO}$ is 140
fs rms [1Hz, 20MHz].

\begin{table}[H]
    \centering
    \begin{tabular}{lcc}
      \toprule
        Carrier feed--through       & -45.4 & dB      \\
        Image frequency suppression & -59.3 & dB      \\
        \midrule
        %MO phase jitter     & 107 fs (rms)  \\
        LO phase jitter             & 140 & fs (rms)  \\
        \midrule
        LO level variation over a week & $< 0.005$ & dBm (p2p)\\
      \bottomrule
    \end{tabular}
    \caption{Measured performances of generated LO}
    \label{table:lo_gen}
\end{table}

\subsubsection{Analog RF Frontend}
\begin{figure}[H]
    \centering
    \includegraphics[angle=90,width=0.9\linewidth]{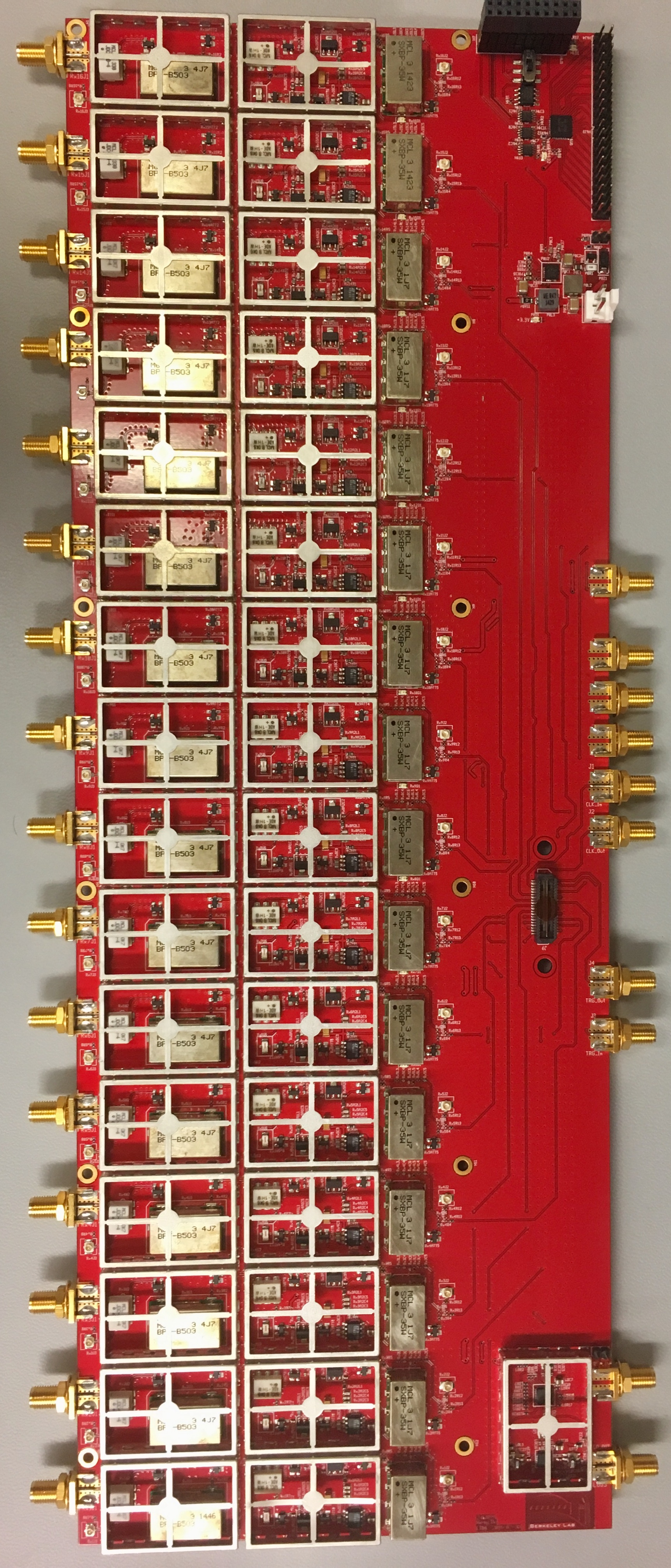}
    \caption{Self--stackable 16 channels down--converter}
    \label{fig:rx16}
\end{figure}

All RF signals are down converted to $f_\text{IF}$ for digitization using a
home-built analog frontend boards. A 16 channel down converter is designed in
order to interface FMC112/FMC116 as shown in Figure \ref{fig:rx16}.

This frontend board features single side band LO generation and distribution,
externally synchronizable switch DC power supply,
individual linear regulator for each channel,
high channel isolations, and
environmental monitoring for voltage, current, LO level and temperature.

The cross talk between channels is optimized by using RF shielding, various
attenuation and filtering in LO path, and individual low noise local voltage
regulators. It is measured as better than 50 dB when using
the high density connector for FMC112/FMC116.
%and $>70$ dB when using separate cables for IF signal output.

\subsubsection{ADC benchmark}

By injecting a near full scale signal at $f_\text{MO}$,
all ADC channels including analog down conversion are benchmarked to match data sheet specifications.
\begin{figure}[H]
    \centering
    \includegraphics[width=.7\linewidth]{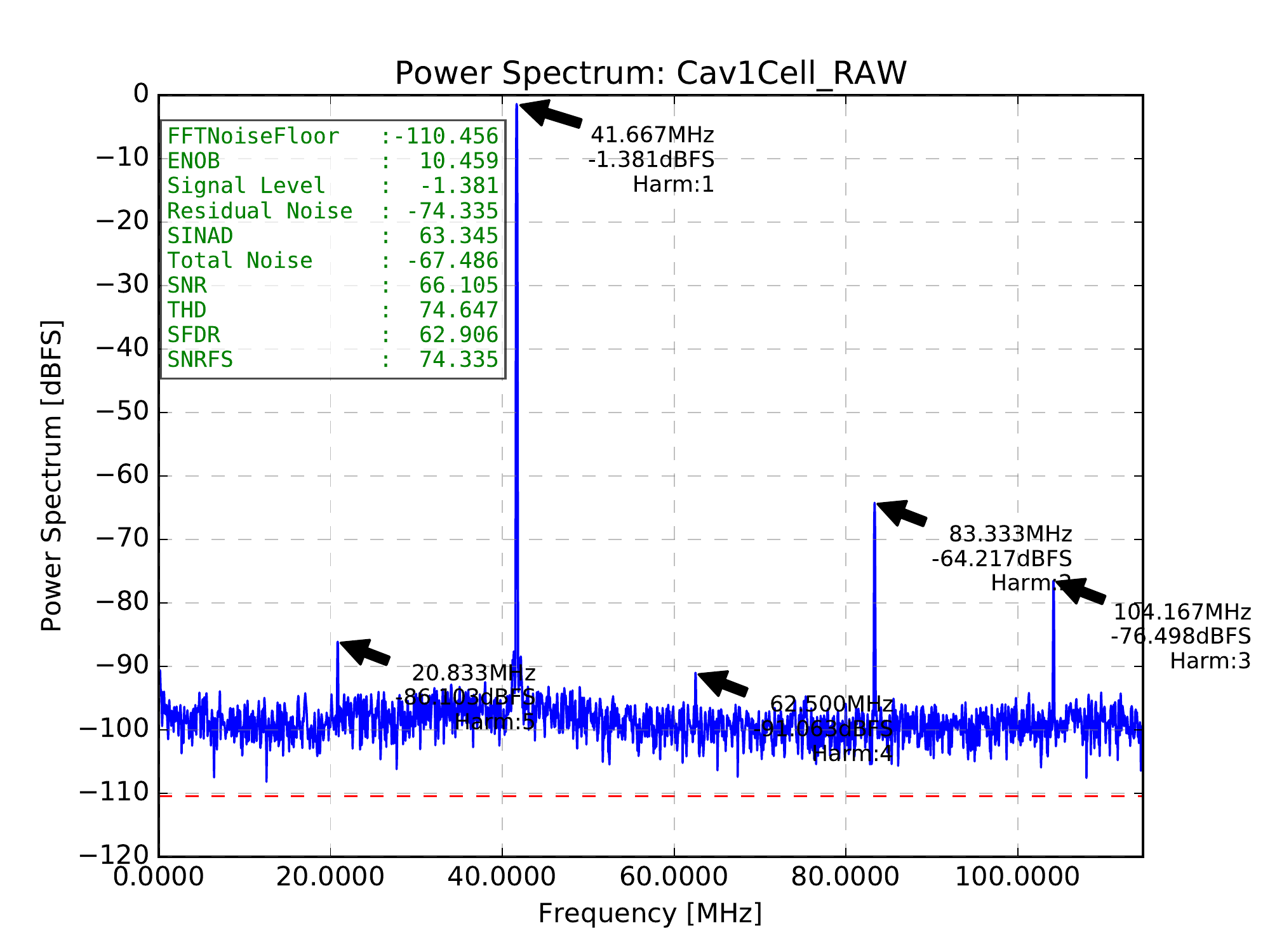}
    \caption{FMC150 ADC1 spectrum}
\end{figure}

In order to separate signal against common components between ADC channels such as $f_\text{MO}$
feedthrough, LO and common power supply noise, we used cross correlation between
channel $x$ and $y$ to measure residual ADC noise \cite{doolittle2016loopback}.

\begin{align*}
    P_\text{res} &= \frac{1}{2}\left( \langle F_x \cdot F_x^\ast \rangle +
    \langle F_y \cdot F_y^\ast \rangle - 2\langle F_x \cdot F_y^\ast
    \rangle\right) \\
    N_\text{res} &= \sqrt{\sum_k P_\text{res}(k)},\qquad 
    \text{SNR}_\text{res} = 20 \log \left( \frac{2^{13}}{\sqrt{2}\cdot N_\text{res}}\right)
\end{align*}

\begin{figure}[H]
  \begin{subfigure}{.5\linewidth}
      \includegraphics[width=\linewidth]{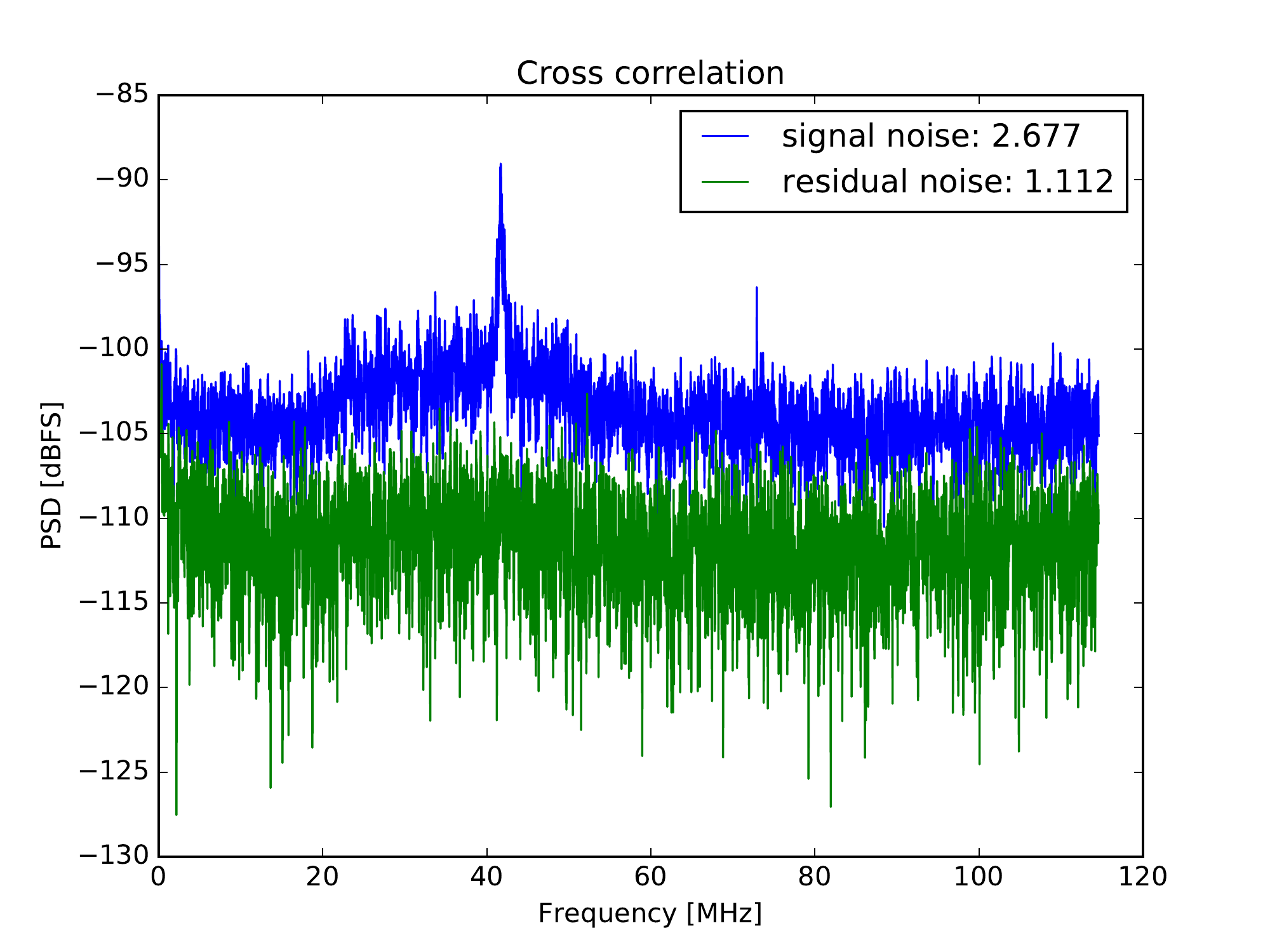}
      \caption{FMC150 residual noise}
  \end{subfigure}
  \begin{subfigure}{.5\linewidth}
      \includegraphics[width=\linewidth]{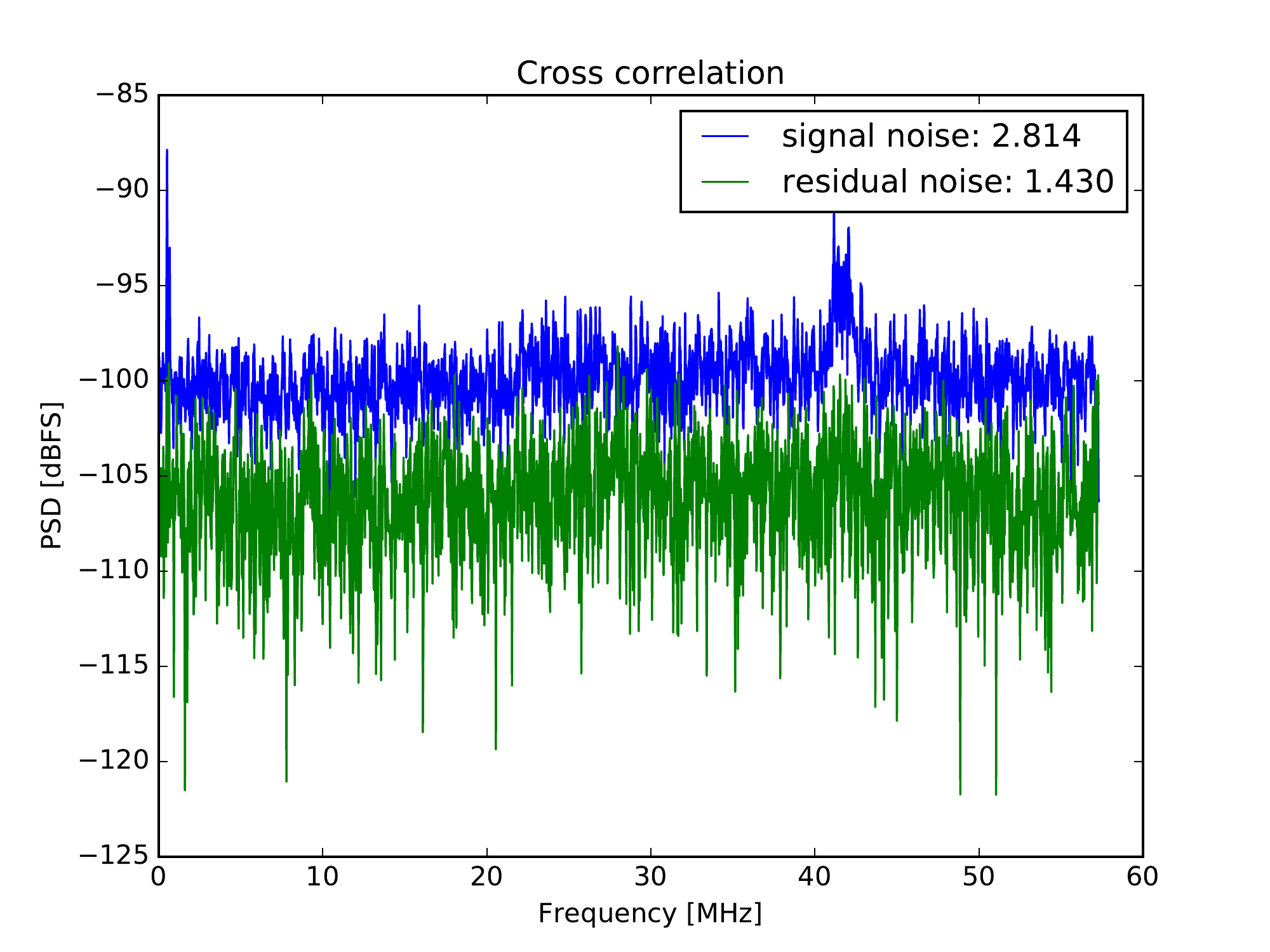}
      \caption{FMC112 residual noise}
  \end{subfigure}
\end{figure}

\begin{table}[H]
    \centering
    \begin{tabular}{lccc}
      \toprule
        & Specified & Measured       & \\
      \midrule
        Total SNR               & 71 & 66 & dBFS \\
        $\text{SNR}_\text{res}$ & 75 & 74.34 & dBFS \\
        Isolation (feedback)    &   & $>72$ & dB   \\
        Isolation (interlock)   & 55 & $>50$ & dB   \\
      \bottomrule
    \end{tabular}
    \caption{Measured ADC benchmarks}
\end{table}

It is concluded that the analog frontend together with digitization platform
meets the design specifications.

\subsubsection{DAC benchmark}
\begin{figure}[H]
  \begin{subfigure}{.5\linewidth}
      \includegraphics[width=\linewidth]{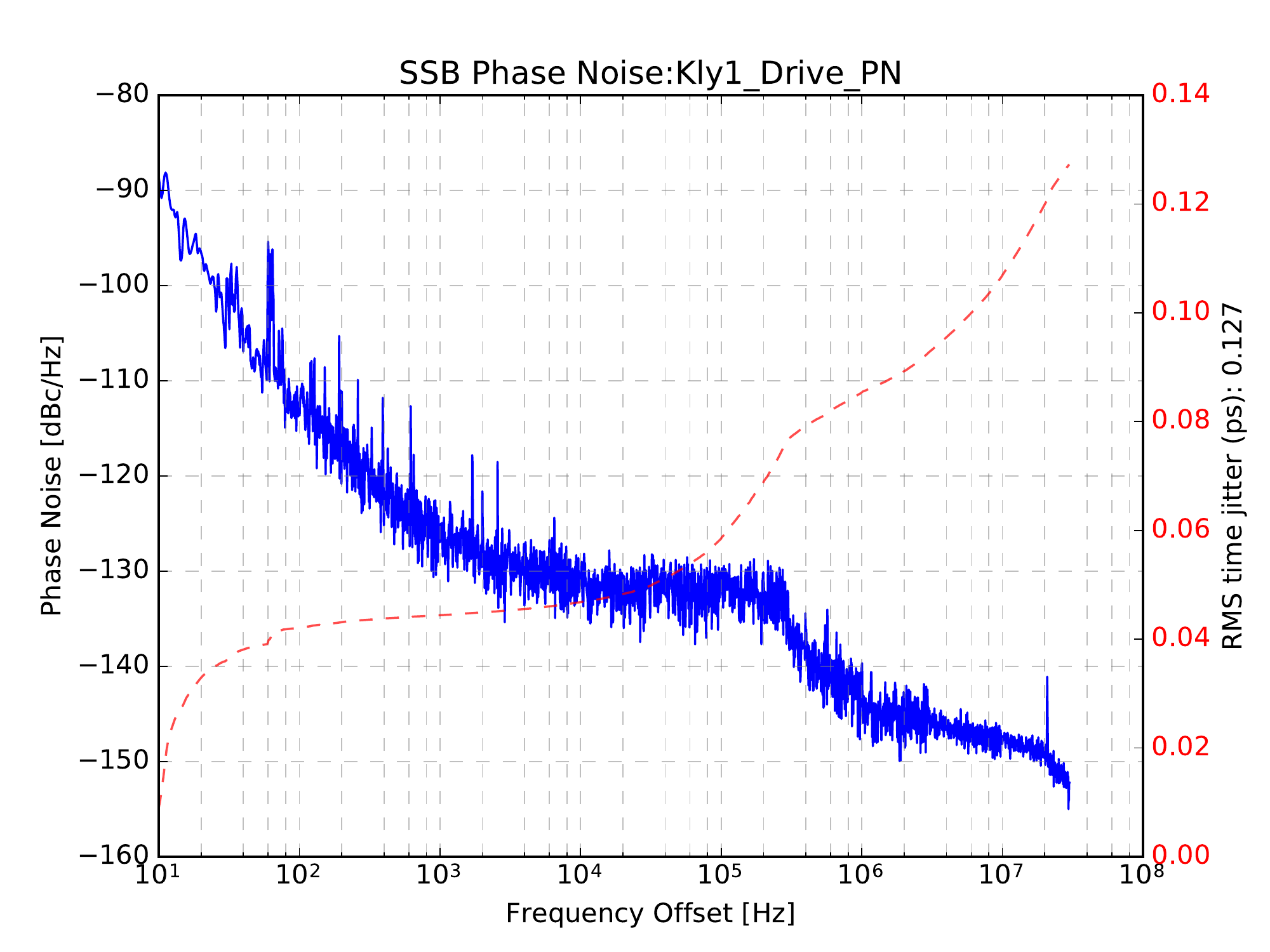}
  \end{subfigure}
  \begin{subfigure}{.5\linewidth}
      \includegraphics[width=\linewidth]{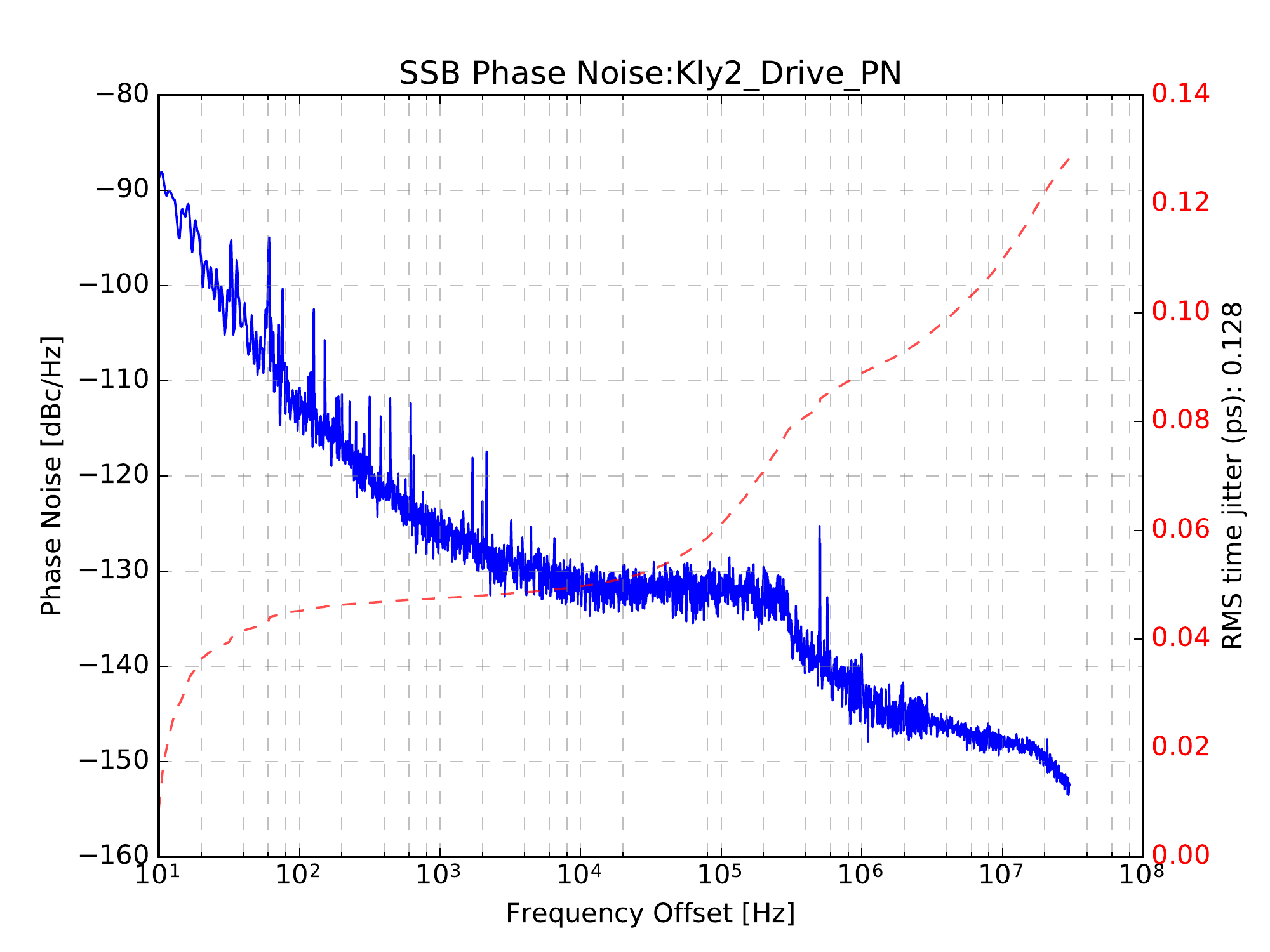}
  \end{subfigure}
    \caption{DAC1 and DAC2 phase noise measurement}
    \label{fig:dac_bench}
\end{figure}

The phase noise of two DAC output signal at $f_\text{MO}$ after up conversion
are measured as $<130$ fs (rms) [1Hz, 20MHz] using a signal source analyzer
(Rohde \& Schwarz FSUP) which also met design specification,
as shown in Figure \ref{fig:dac_bench}.

\subsection{RF Monitor chassis}

\begin{figure}[H]
    \centering
    \includegraphics[width=\linewidth]{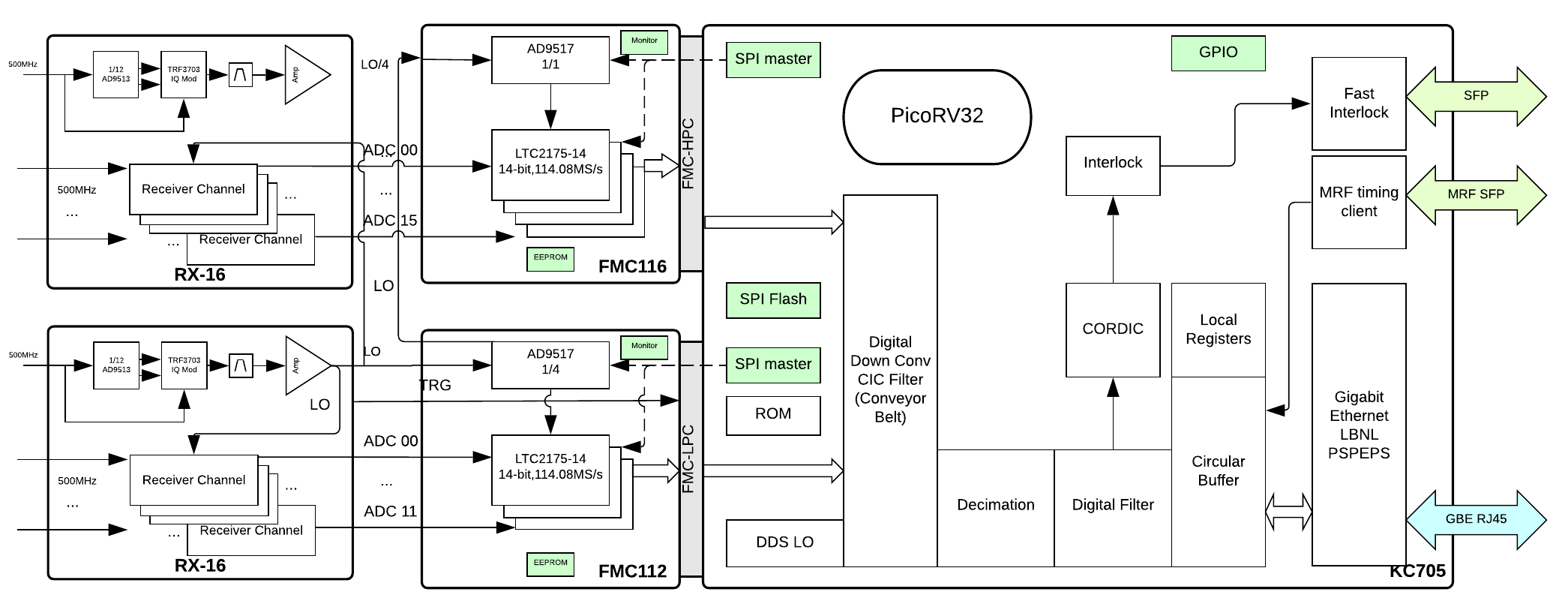}
    \caption{RF Monitor chassis hardware configuration}
    \label{fig:rfmon_hardware}
\end{figure}

The RF monitor chassis uses Abaco Systems FMC112 and
FMC116 with Xilinx KC705 FPGA carrier, as in Figure \ref{fig:rfmon_hardware}.
Two 16--channel down conversion frontend boards are used to host 28 channels.
LO signal are from LLRF chassis to keep the same frequency divider states for
RF phase measurement. All sampling frequencies are at $f_\text{S2}$ and FPGA
clock is running at twice as fast at $f_\text{S1}$.
The communication interfaces are common to LLRF chassis.

\subsection{Fast Interlock Chassis}
Also based on Xilinx KC705, the Fast Interlock Chassis uses two GTX links to
collect and process IQ streams from LLRF and RFMON chassis, with addition of 16 channels of
arc detectors faults and arc power faults through a Field IO FPGA (Digilent CMOD S6).
The standalone reference clock is tuned to be within $\pm 100$ ppm of stream
clock at $f_\text{dsp}$.

\begin{figure}[H]
    \centering
    \includegraphics[width=0.7\linewidth]{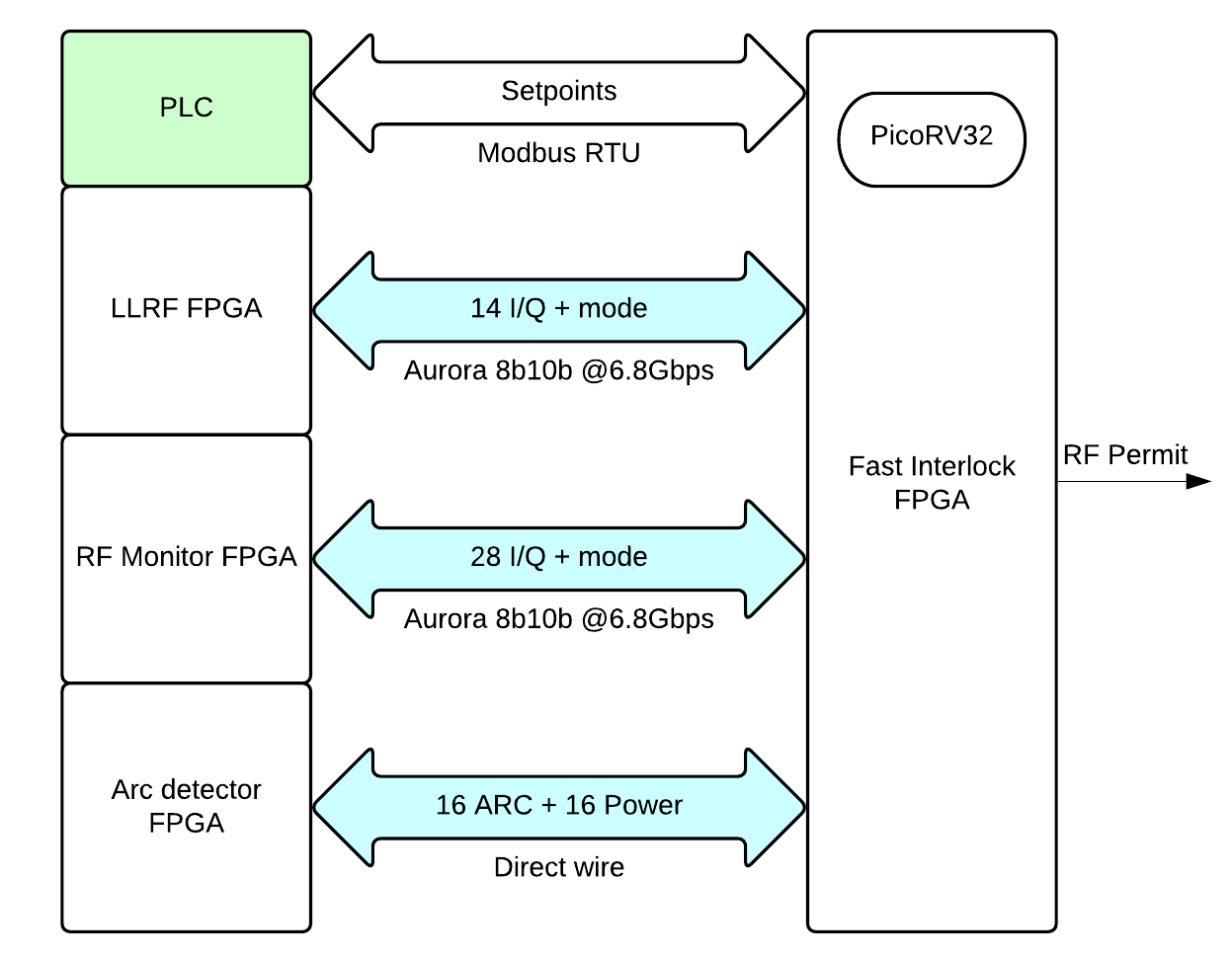}
    \caption{Fast Interlock Chassis logic diagram}
\end{figure}

All interlock configurations and setpoints are passed by PLC--FPGA communication, and the
end--to--end interlock latency needs to be $<4 \mu$s.

\section{Firmware design}
\label{sec:firmware}

\subsection{Digital Signal processing}

\subsubsection{Direct Digital Down-conversion}

Non--IQ sampling avoids aliasing for high precision digitization.\cite{doolittle2006digital}
In our case, with $\tfrac{f_\text{IF}}{f_\text{Sample}} = \frac{2}{11} = \theta
\simeq 65.45^\circ$,
one can construct direct digital down--conversion (DDC) using two consecutive ADC samples
$y_n, y_{n+1}$. The coefficient matrix is generated using a Coordinate Rotation
Digital Computer (CORDIC).
\begin{align*}
    \begin{pmatrix}
        y_n \\
        y_{n+1}
    \end{pmatrix}
    &=
    \begin{pmatrix}
        \cos(n\theta)       & \sin(n\theta) \\
        \cos((n+1)\theta)   & \sin((n+1)\theta)
    \end{pmatrix}
    \begin{pmatrix}
        I \\
        Q
    \end{pmatrix} \\
    \begin{pmatrix}
        I \\
        Q
    \end{pmatrix}
    &= \frac{1}{\sin\theta}
    \begin{pmatrix}
        \sin((n+1)\theta)    & -\sin(n\theta) \\
        -\cos((n+1)\theta)   & \cos(n\theta)
    \end{pmatrix}
    \begin{pmatrix}
        y_n \\
        y_{n+1}
    \end{pmatrix}
\end{align*}

\subsubsection{Double-time signal process}

\begin{figure}[H]
    \includegraphics[width=\linewidth]{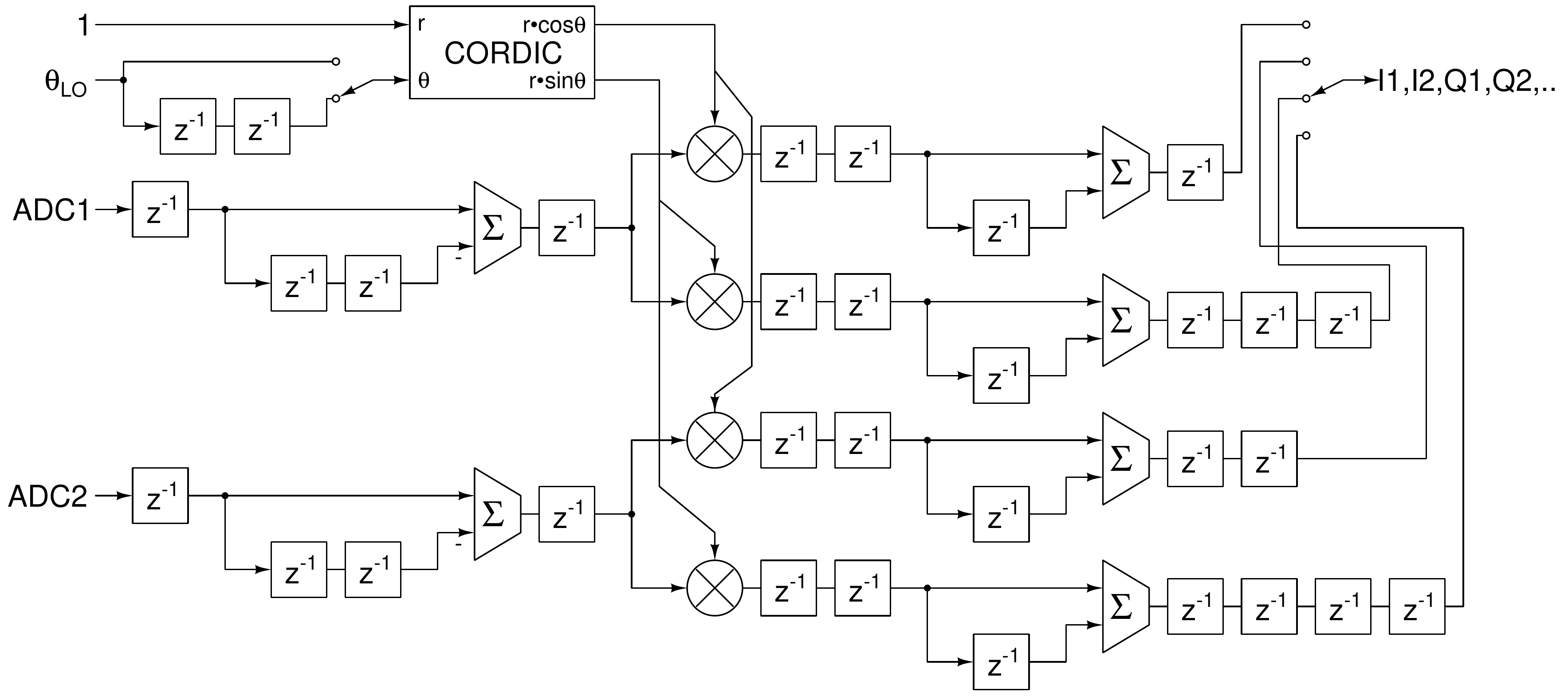}
    \caption{double--time: Two DSP clock cycles per ADC sample}
    \label{fig:doubletime}
\end{figure}

Because $f_\text{dsp} = 2f_\text{S2}$, and also the need for dynamically
configure feedback signal paths according to RF drive mode, it is desired to
have two DSP clock cycles per ADC sample in order to reuse the same pipeline for
different combinations of RF signal processing. Figure \ref{fig:doubletime}
shows an example of pipe lining two ADC channels to a stream of digitally down
converted IQ pairs.

\subsubsection{Feedback controller}

\begin{figure}[H]
    \centering
    \includegraphics[width=\linewidth]{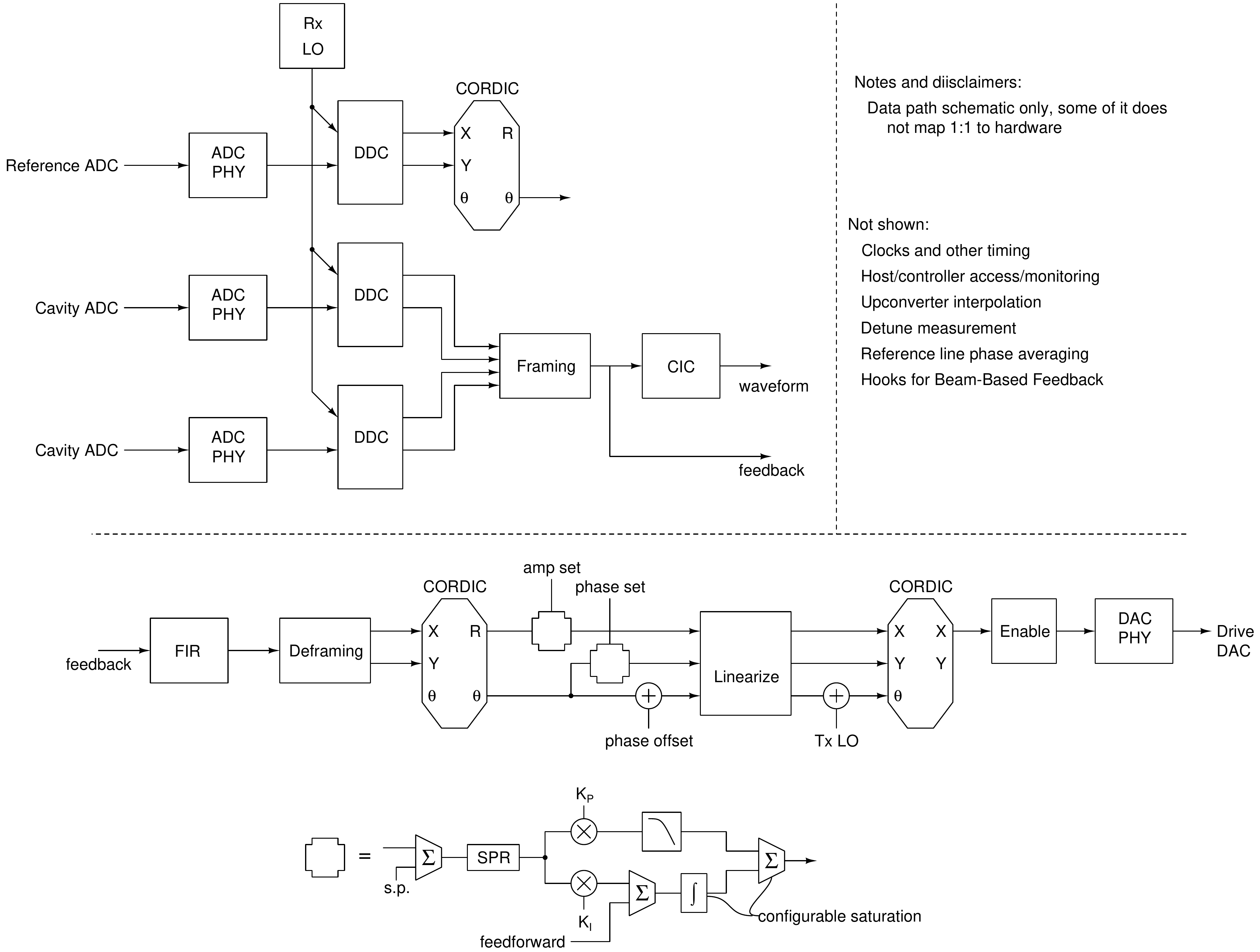}
    \caption{LLRF firmware overview}
    \label{fig:loops}
\end{figure}

The feedback controller is constructed as shown in Figure \ref{fig:loops}.

The framed IQ pairs, in our case two cavity probe signals are converted to
amplitude and phase in order to compare with loop sepoints, and fed into a set
of proportional--integral controllers with limiter. The clipped output is then
digitally up converted using an output CORDIC with an optional phase shifter to
compensate variable loop group delay. When RF permit signal is valid, the loop output
signal is sent to each DAC.

There are amplitude and phase loops for each klystron. For one klystron driving
two cavities of test loads, the weighted average of two cavity probe amplitudes
are used for amplitude loop, and one cavity phase for phase loop.

\[
    C(z) = K_p + K_i \frac{1}{1-z^{-1}}, \qquad T=\frac{4}{f_\text{clk}}
\]

Since the exact controller transfer function is known by design, the loop
register scaling are verified using an IIR cavity emulator in $z$ domain.

\subsubsection{Waveform}

All baseband signal after DDC are serialized onto a conveyor belt signal stream
to apply a run--time configurable CIC filter for different decimation factors.

\begin{figure}[H]
    \centering
    \includegraphics[width=0.6\linewidth]{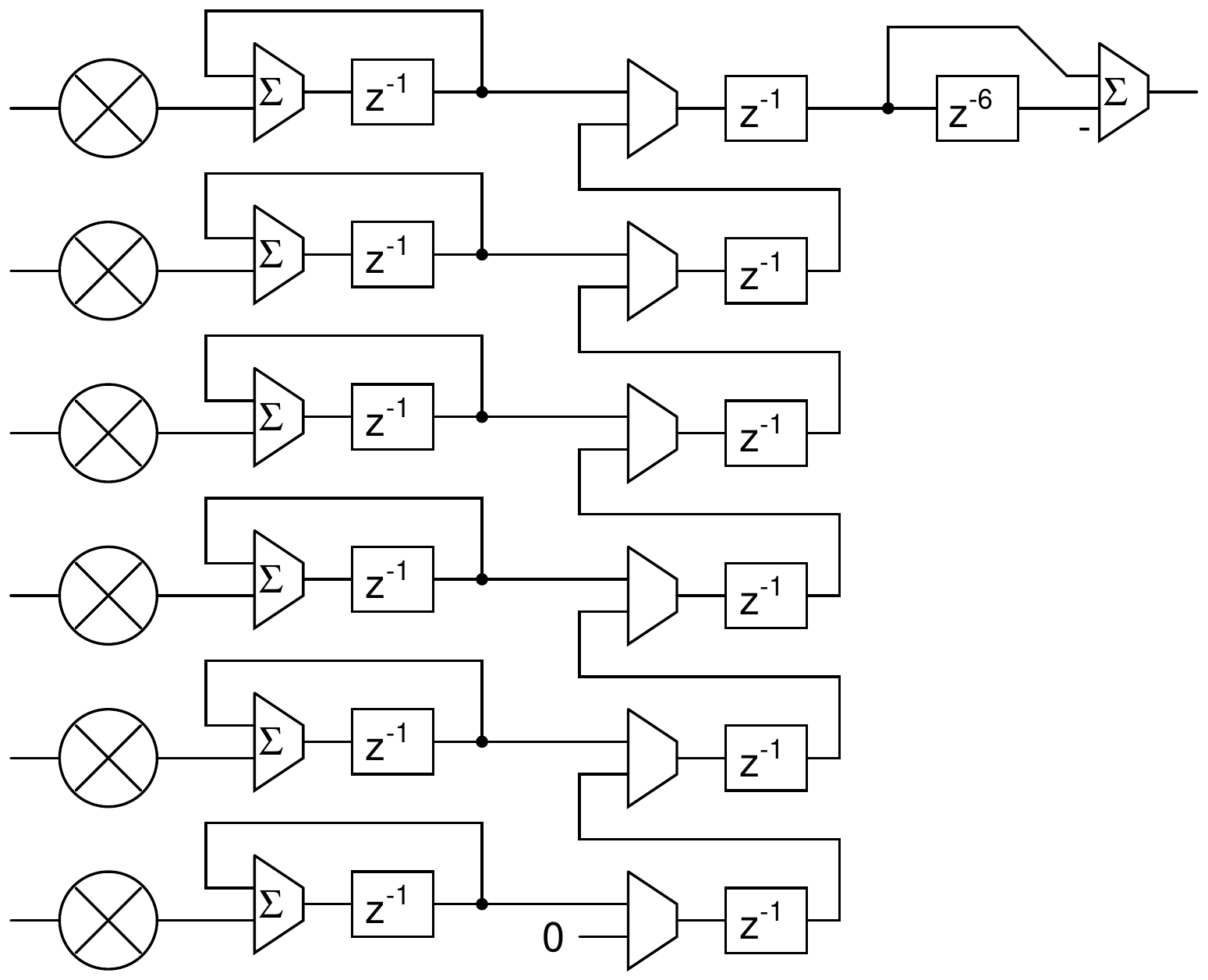}
    \caption{signal pipe line with CIC filter}
\end{figure}

Features of waveform handling includes dynamic channel selection,
collision--free doubled buffer structure,
associated statistics with timestamps,
and fault capturing.

\begin{figure}[H]
  \begin{subfigure}{.5\linewidth}
      \includegraphics[width=\linewidth, page=2]{fig/fault_waveforms.pdf}
      \caption{Cavity forward signals}
  \end{subfigure}
  \begin{subfigure}{.5\linewidth}
      \includegraphics[width=\linewidth, page=1]{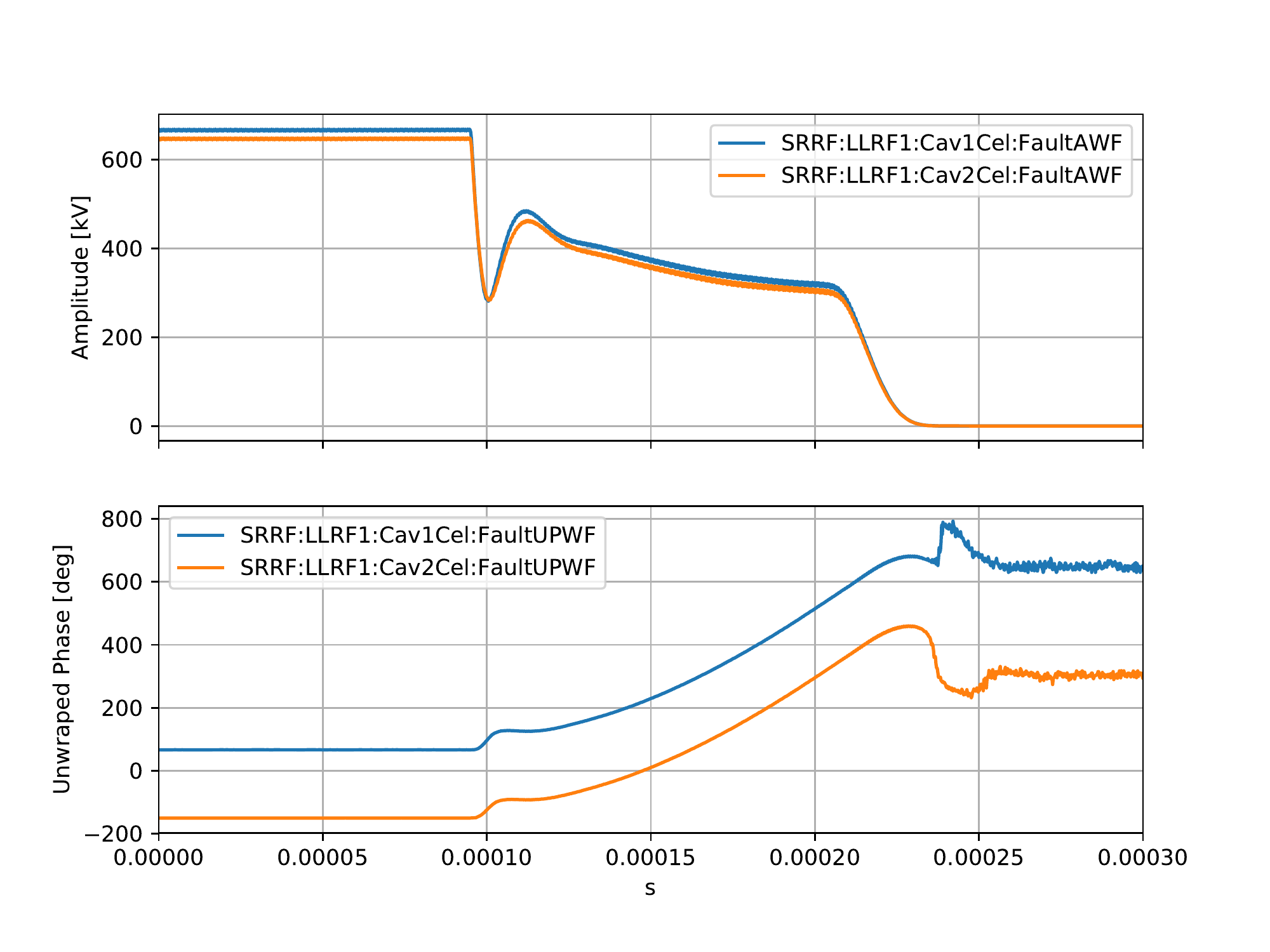}
      \caption{Cavity probe signals}
  \end{subfigure}
  \caption{Fault Waveform Capturing}
    \label{fig:fault}
\end{figure}

Figure \ref{fig:fault} shows an example of fault waveform capturing
at an event of RF trip, when both cavities were only driven by beam
after LLRF permit was removed.
This is an essential feature for finding the root cause of RF system trip.

\subsubsection{Peripheral management}
There are hundreds of registers and complicated processes involved in
initialization process to bring the system to nominal state from power up.
An open source RISC--V soft core \texttt{PicoRV32}\cite{clifford2015picorv32}
is used in each design to handle peripheral management and initialization
process.
\begin{figure}[H]
    \centering
    \includegraphics[width=0.8\linewidth]{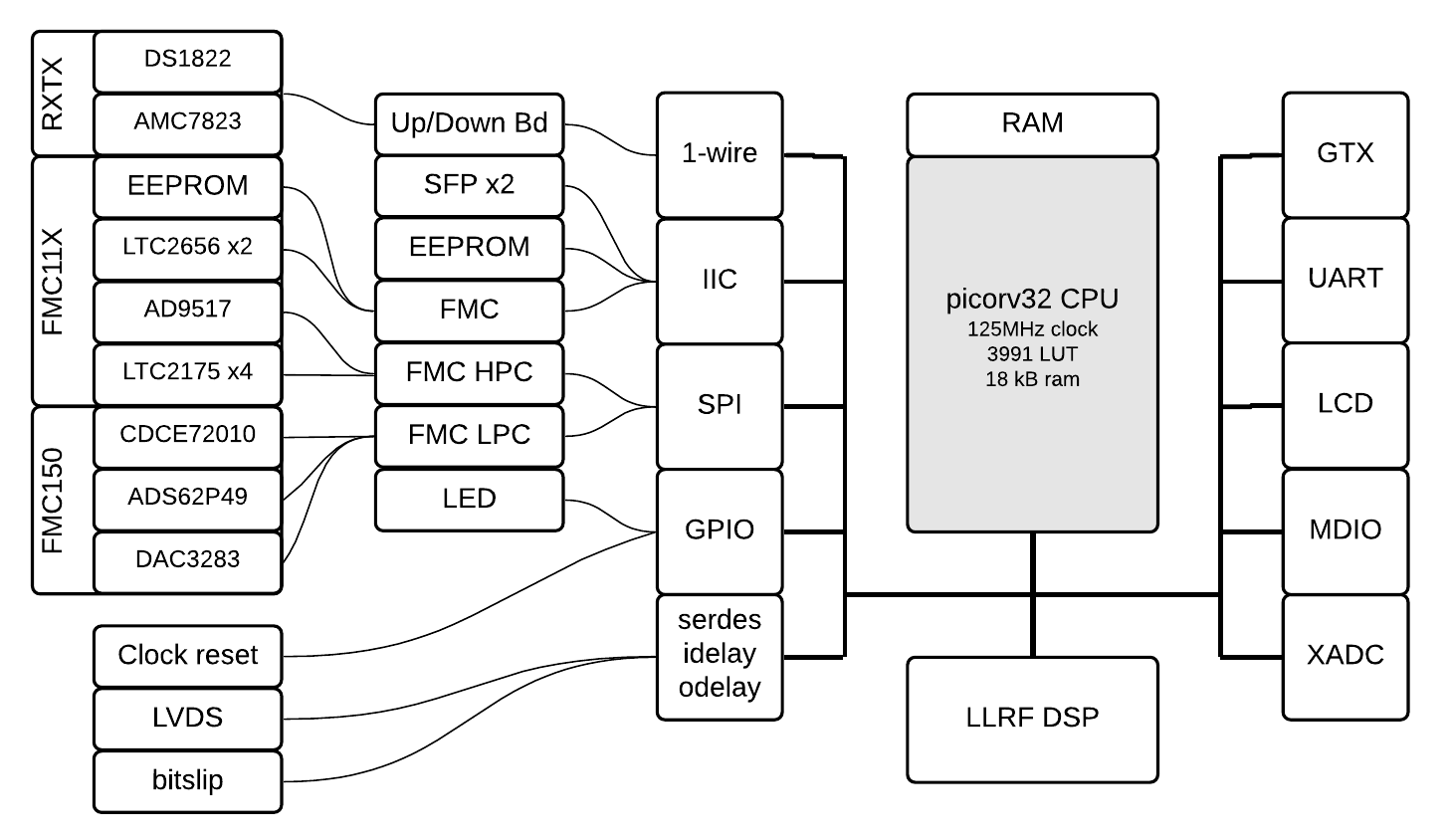}
    \caption{Peripheral management and power--on initialization by CPU}
\end{figure}

\subsubsection{PLC--FPGA interface}
\begin{figure}[H]
    \centering
    \includegraphics[width=0.6\linewidth]{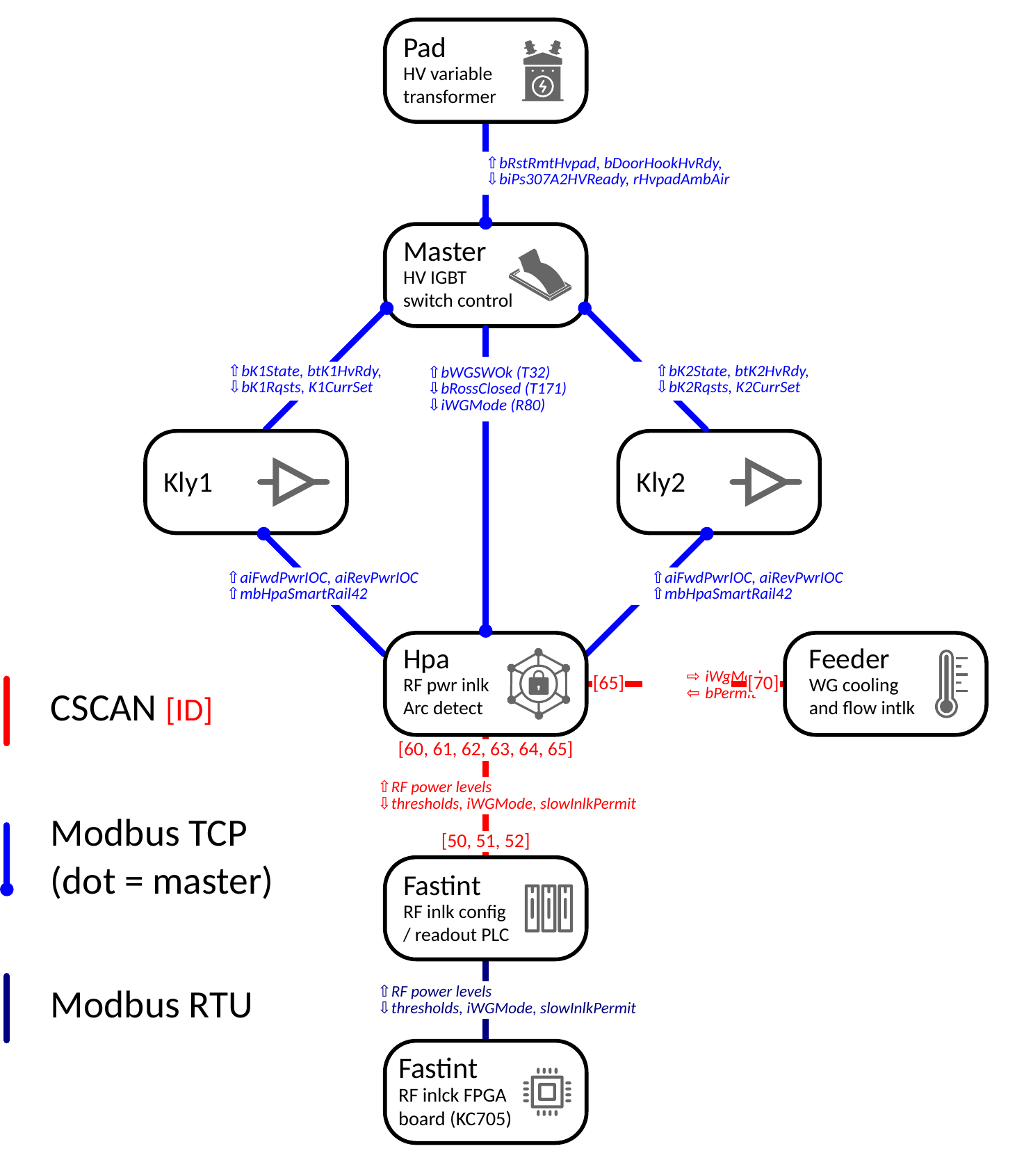}
    \caption{ALS SRRF PLC--FPGA interfaces}
    \label{fig:plc}
\end{figure}

\texttt{PicoRV32} in Fast Interlock Chassis also handles direct interfacing with PLC via
\texttt{ModbusRTU}. The multi-chassis FPGA system works as a peripheral from
PLC point of view as shown in Figure \ref{fig:plc}.

\section{Software design}
\label{sec:software}
Both python based diagnostic tools and EPICS based operation applications were developed
based on the UDP Gigabit direct access to registers and waveforms
\cite{serrano2011fpga}.

\section{Operation and performance}
\label{sec:performance}

\subsection{RF stability}
\begin{figure}[!t]
  \begin{subfigure}{0.5\linewidth}
      \includegraphics[width=\linewidth]{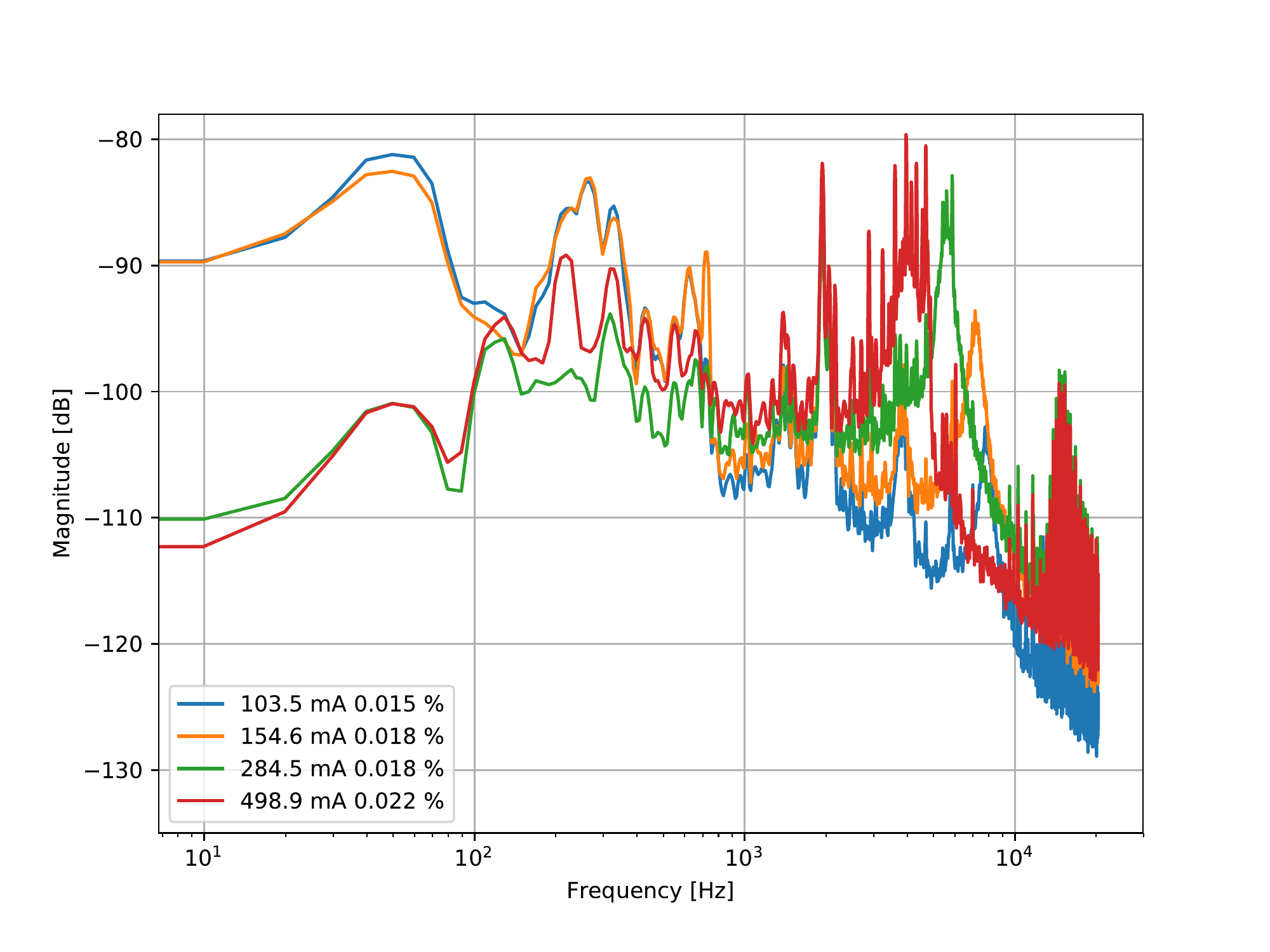}
      \caption{Amplitude loop}
  \end{subfigure}
  \begin{subfigure}{0.5\linewidth}
      \includegraphics[width=\linewidth]{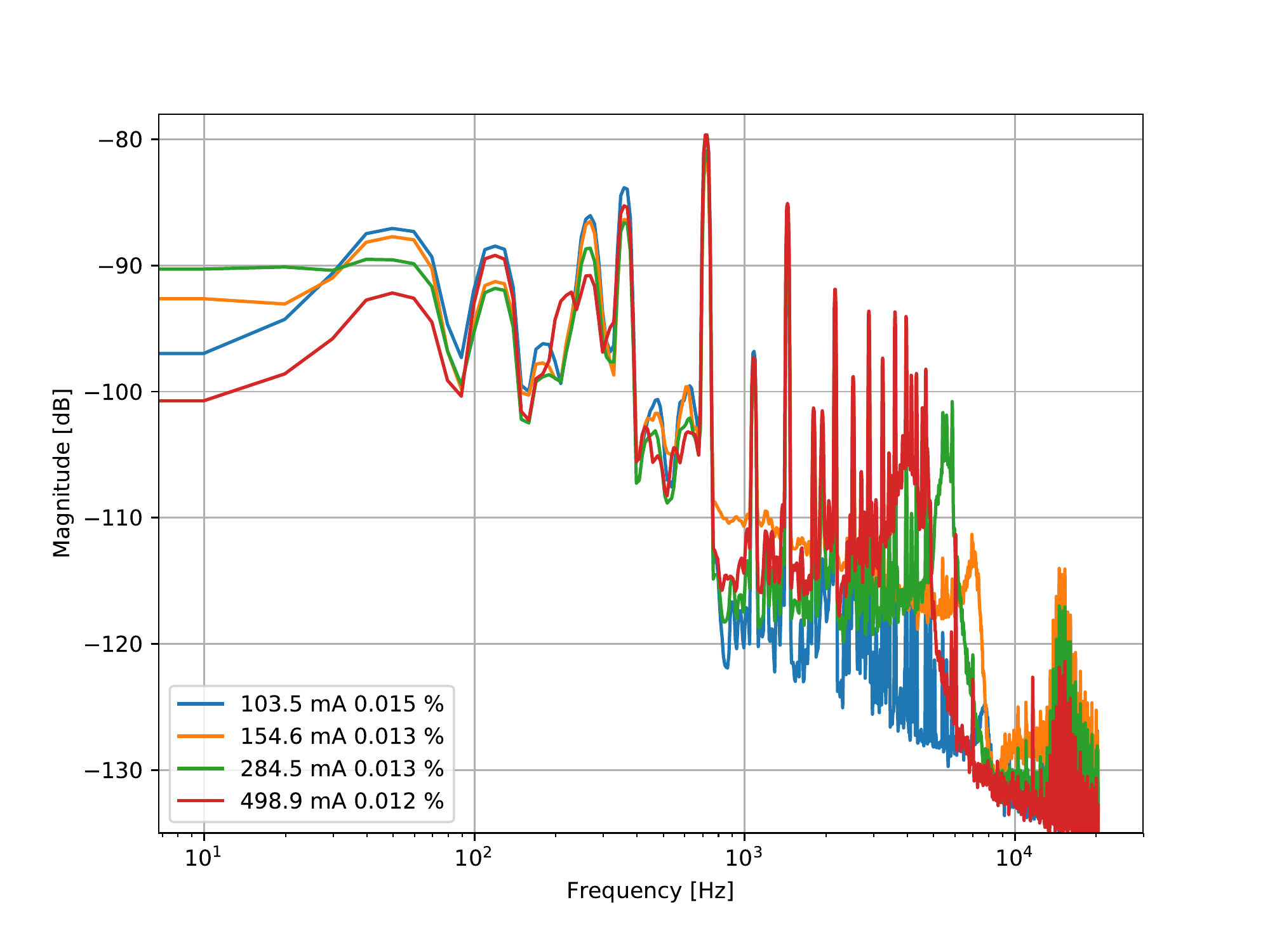}
      \caption{Phase loop}
  \end{subfigure}
    \caption{LLRF loop noise spectrum density}
    \label{fig:loop_performance}
\end{figure}

We have measured both amplitude and phase loop noise spectrum density from in--loop waveform data
at different beam currents as shown in Figure \ref{fig:loop_performance}. 
The data was measured under klystron 2 drive mode, where average of two cavity cell voltages
are regulated.  Loop parameters are at operation nominal setting,
with $\sim 1$ kHz bandwidth, integral only.
It is observed that,
as beam current increased from 103 mA to 499 mA, 
the synchrotron frequency moved from $<10$ kHz towards $\sim 4$ kHz,
and its contribution became more significant.
Within analyse bandwidth of [10Hz, 20kHz], the amplitude loop stability is
measured as $< 0.03\%$ rms, and phase loop stability is $<0.02^\circ$ rms.

\subsection{Loop Frequency Response}

\begin{figure}[H]
  \begin{subfigure}{0.5\linewidth}
      \includegraphics[width=\linewidth]{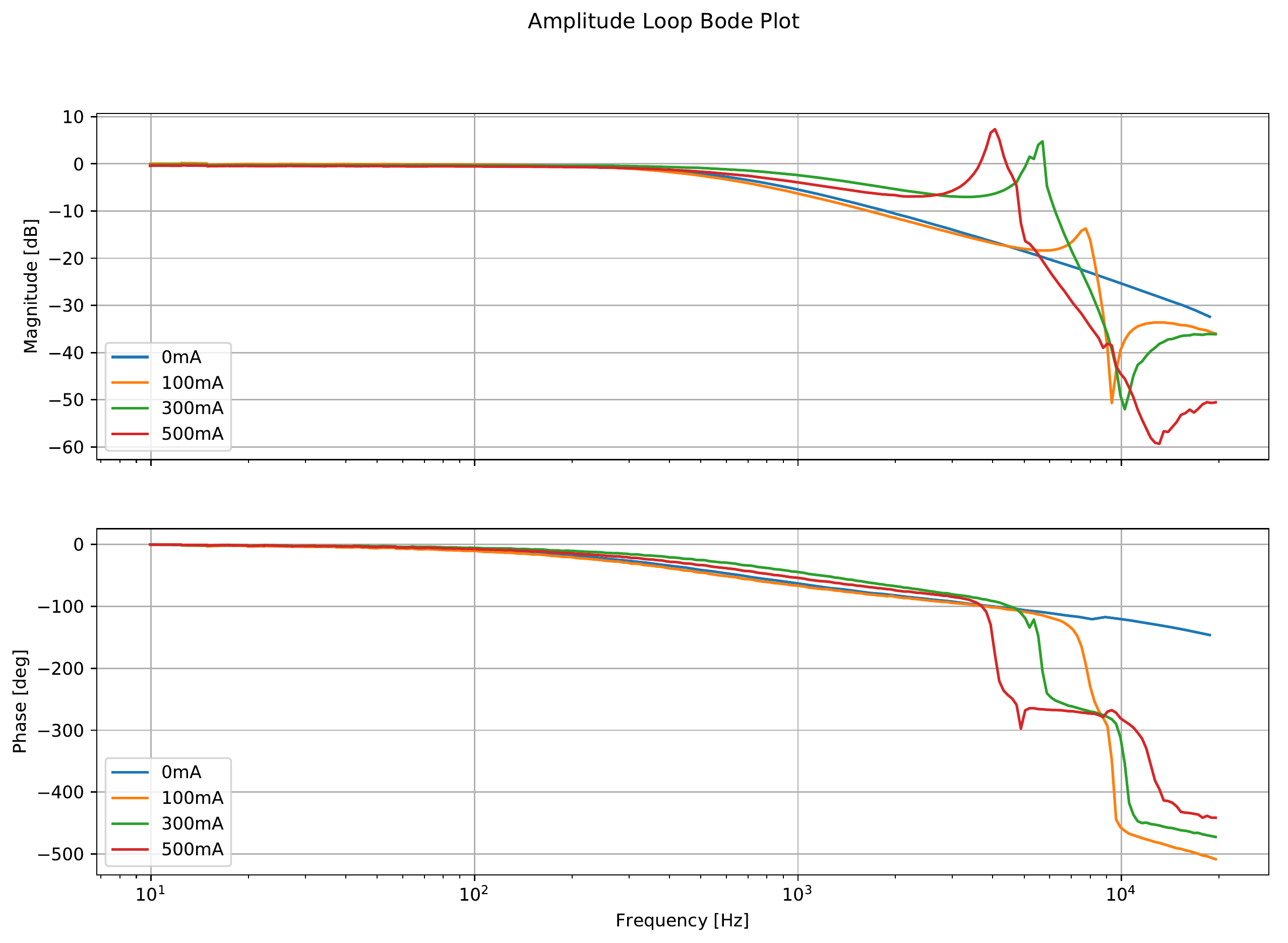}
      \caption{Amplitude loop}
  \end{subfigure}
  \begin{subfigure}{0.5\linewidth}
      \includegraphics[width=\linewidth]{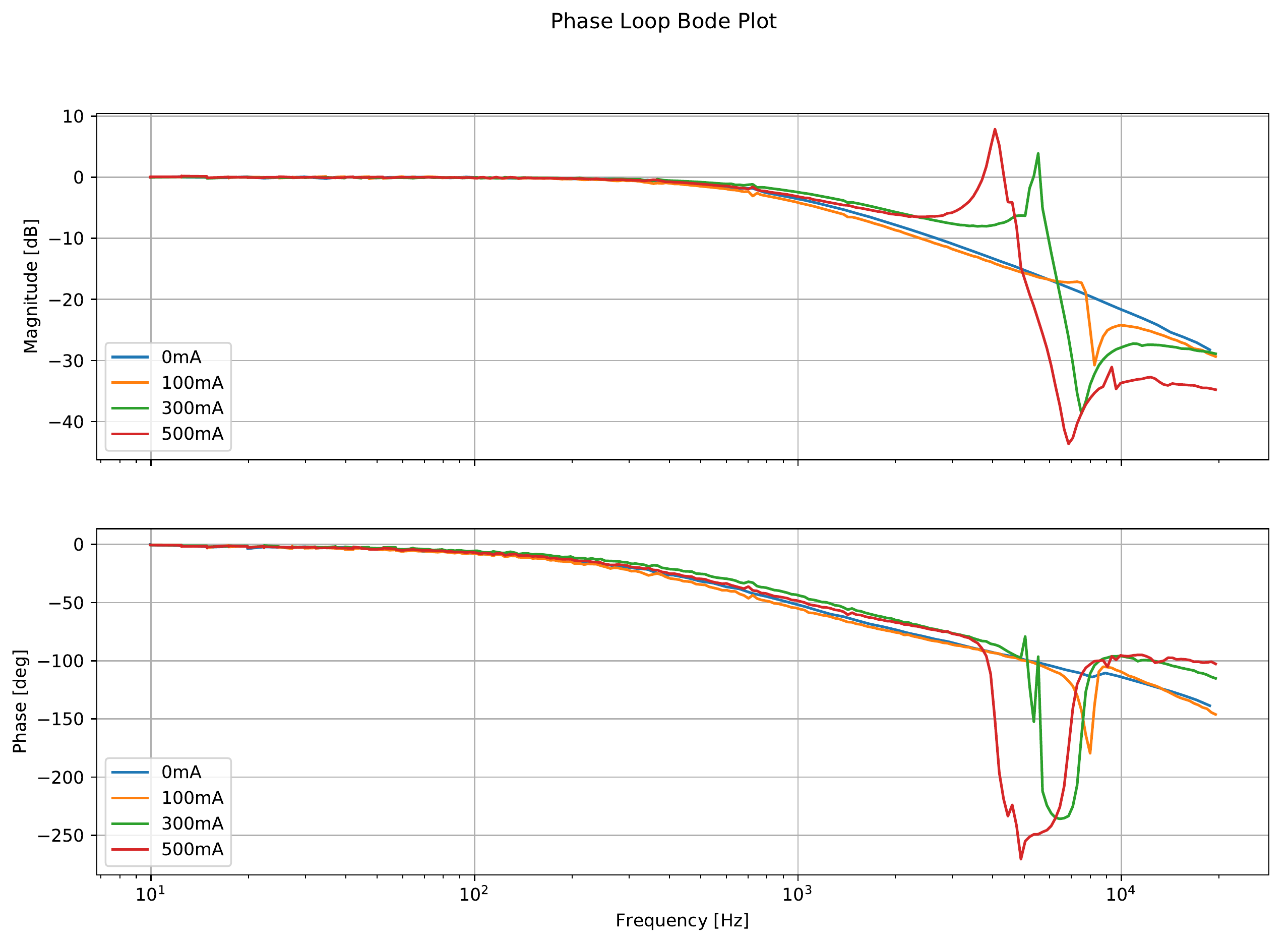}
      \caption{Phase loop}
  \end{subfigure}
  % \begin{subfigure}{0.5\linewidth}
  %     \includegraphics[width=\linewidth]{fig/am2pm.pdf}
  %     \caption{Beam induced amplitude to phase modulation}
  % \end{subfigure}
  % \begin{subfigure}{0.5\linewidth}
  %     \includegraphics[width=\linewidth]{fig/pm2am.pdf}
  %     \caption{Beam induced phase to amplitude modulation}
  % \end{subfigure}
  \caption{LLRF loop frequency responses}
  \label{fig:loop_bode}
\end{figure}

By injecting a excitation tone on either amplitude or phase loop setpoint, the
closed loop frequency response could be measured as a function of excitation
frequency. We have measured bode plots at different beam currents as shown in
Figure \ref{fig:loop_bode}. Both amplitude and phase loop unity gain is observed
around 1 kHz, as expected. Again, beam induced synchrotron frequencies
contributes as a resonance peak outside loop bandwidth.

\subsection{Hard real--time RF interlock}
\begin{table}[H]
    \centering
    \begin{tabular}{lcc}
      \toprule
        RF Power (Lab)     & 1.45  & $\mu$s  \\
        RF Power (ALS)     & $< 3$ & $\mu$s  \\
        ARC det. latency   & $< 2$ & $\mu$s  \\
      \bottomrule
    \end{tabular}
    \caption{End--to--end measured RF interlock latency}
\end{table}

\section{Conclusion}
\label{sec:conclusion}
The digital LLRF system for ALS storage ring RF is operational since March 2017
with $>6$ months of mean time between failure.
System functionality and performance met design requirements.

\section*{Acknowledgment}

This work is supported by the Office of Science, Office of Basic Energy Sciences, of the U.S.
Department of Energy under Contract No. DE-AC02-05CH11231.

\bibliographystyle{IEEEtran}
\bibliography{IEEEabrv,reference}

\end{document}